\documentclass[twocolumn,aps,prl,showpacs,amsmath,amssymb,nobalancelastpage]{revtex4}
\usepackage{empheq}
\usepackage[dvips]{graphicx}
\usepackage{bm,eucal}
\usepackage{delarray,fancybox}
\usepackage{mathdots}
\newcommand{\dif}{\mathrm{d}}
\newcommand{\dt}{\mathrm{d}t}
\newcommand{\ti}{t_{\mathrm{o}}}
\newcommand{\tf}{t_{\mathrm{f}}}
\newcommand{\str}[1]{\overset{\diamond}{#1}}
\newcommand{\oti}[1]{\overset{\triangleright}{#1}}
\begin{document}
\title{On the use of stochastic differential geometry for non-equilibrium thermodynamics modeling and control}

\author{Paolo Muratore-Ginanneschi}
\affiliation{Department of Mathematics and Statistics, 
University of Helsinki PL 68, 00014 Helsinki, Finland.}
%\email{paolo.muratore-ginanneschi@helsinki.fi}
%\date{\today}
%

\begin{abstract}
We discuss the relevance of geometric concepts in the theory of stochastic differential equations
for applications to the theory of non-equilibrium thermodynamics of small systems. In particular,
we show how the Eells-Elworthy-Malliavin covariant construction of the Wiener process on a Riemann 
manifold provides a physically transparent formulation of optimal control problems of finite-time 
thermodynamic transitions. Based on this formulation, we turn to an evaluative discussion of recent 
results on optimal thermodynamic control and their interpretation.  
\end{abstract}
\maketitle

\section{Introduction}

Stochastic differential equations \cite{IkedaWatanabe} provide a widely applied mathematical 
model for non-equilibrium dynamics. In particular, they are well adapted to the description of
kinetics and finite time thermodynamics of small systems such as bio-molecules, RNA and other molecular 
scale ``machines'', see e.g. \cite{Rit08} %and references therein 
for review of theoretical and experimental aspects. 
The application of stochastic differential equations to non-equilibrium phenomena bears a natural 
relation to geometry in an, at least, twofold way. First, even when a stochastic differential 
equation is used to model processes evolving in a flat, Euclidean, space, 
in \emph{arbitrary coordinates} the scale of the noise in the stochastic differential equation 
imposes a Riemannian or sub-Riemannian (if there are constraints on the admissible directions 
of motion such as conditioning a set of variables to be the derivatives of other ones) metric 
on the space. The second way geometry sets in is more
%a second source of geometrical concepts more 
distinctive of applications to non-equilibrium physics. It originates from the use of second 
derivatives of thermodynamic potentials to impose a Riemannian metric on the set 
of equilibrium states. The consequent notion of thermodynamic length has been then exploited 
to characterize optimal processes in macroscopic and, very recently, microscopic, 
nano-scale processes (see \cite{SiCr12} for references and discussion).
In the present contribution we show how a coordinate-independent geometric formalism may ease the
analysis of control problems arising in the study of finite-time thermodynamics of small systems. 
Although the needed mathematical tools are well known in stochastic analysis 
(see e.g.\cite{IkedaWatanabe,Nor92,Hsu02,Baudoin}), we are not aware of any previous application to
stochastic thermodynamics. %Thus, it seems to us that a concise discussion of their immediate 
%implications for small systems non-equilibrium thermodynamics may be a service to the community. 
In what follows we will generically denote by $\mathbb{M}$ a complete connected Riemannian manifold. 
We will also imply some further technical assumptions to guarantee probability 
conservation (see e.g. \cite{Hsu08} for concise discussion).
In practice, our main focus here will be on covariance of physical laws
so that $\mathbb{M}$ can be thought as $\mathbb{R}^{d}$ endowed with a Riemannian metric. Finally, in order to simplify the
notation, thermodynamic expressions will be evaluated at unit temperature.

\section{Geometry and the development map}

Differential geometry is most naturally formulated in the language of first order forms. For this reason
Stratonovich calculus \cite{IkedaWatanabe} is commonly considered to have an edge on Ito calculus when 
it comes to identify geometric structures. Let us therefore start by considering a diffusion process
 $\boldsymbol{\xi}\equiv\left\{\boldsymbol{\xi}_{t}; t\in [\ti,\tf]\right\}$ specified by stochastic 
differential equations in the sense of Stratonovich %of the form 
\begin{eqnarray}
\label{sde:Stratonovich}
\dif\boldsymbol{\xi}_{t}=\boldsymbol{b}_{S}(\boldsymbol{\xi}_{t},t)\,\dt
+\boldsymbol{e}_{i}(\boldsymbol{\xi}_{t})\str{}\dif \beta_{i;t}
\end{eqnarray}
in a coordinate neighborhood $\mathbb{U}\in\mathbb{M}$ centered e.g.
around the initial condition $\boldsymbol{\xi}_{\ti}=\boldsymbol{x}_{o}$. The symbol $\str{\,}$ pinpoints the Stratonovich
prescription. 
In (\ref{sde:Stratonovich}) $\boldsymbol{b}_{S}\colon \mathbb{U}\times [\ti,\tf]\mapsto \mathbb{R}^{d}$
is some smooth \emph{map}, Einstein convention is implied on the latin 
indices $i=1,\dots,d$ to pair the elements of a collection of $d$ independent one-dimensional Euclidean 
Wiener processes (Brownian motions)  
$\beta_{i}\equiv\left\{\beta_{i;t}; t\in [\ti,\tf]\right\}$ 
to $d$ orthonormal frame fields $\left\{\boldsymbol{e}_{i}\right\}_{i=1}^{d}$ specifying a basis for $\mathrm{T}_{\boldsymbol{\xi}_{t}}\mathbb{M}$. 
Let us also endow $\mathbb{M}$ with a strictly positive, time-independent metric tensor 
$\mathsf{g}$ and require that for any $\boldsymbol{x}\in\mathbb{U}$
\begin{eqnarray}
\label{}
\langle\boldsymbol{e}_{i}\,,\boldsymbol{e}_{j}\rangle_{\mathsf{g}}:= \mathsf{g}_{\alpha_{1}\,\alpha_{2}} 
\boldsymbol{e}_{i}^{\alpha_{1}}\boldsymbol{e}_{j}^{\alpha_{2}}=\delta_{ij}
\end{eqnarray}
($\alpha_{j}=1,\dots,d$, $j=1,2$) which is equivalent to
\begin{eqnarray}
\label{}
\boldsymbol{e}_{i}\otimes\boldsymbol{e}_{i}= \mathsf{g}^{-1}
\end{eqnarray}
The hypothesis of time-independence of the metric tensor is physically relevant and will also serve 
later to neaten the formalism. The extension to the time dependent case is, however, straightforward 
and it is of common use for example in the inquiry of Ricci flows, see e.g. \cite{Topping,Cou11} 
for details. From (\ref{sde:Stratonovich}) we can write for any $\boldsymbol{x}\in\mathbb{U}$ the generator
of the stochastic process $\boldsymbol{x}$ acting on any \emph{scalar function} 
$f:\mathbb{U}\mapsto \mathbb{R}$
\begin{eqnarray}
\label{gen:Stratonovich}
\lefteqn{
\mathfrak{L}_{S}f=\left(\boldsymbol{b}_{S}\cdot\partial_{\boldsymbol{x}}
+\frac{1}{2}\boldsymbol{e}\cdot\partial_{\boldsymbol{x}}
\boldsymbol{e}\cdot\partial_{\boldsymbol{x}}\right)\,f
=}
\nonumber\\&&
\hspace{-0.2cm}
\left\{\left[\boldsymbol{b}_{S}+\frac{1}{2}(\boldsymbol{e}_{i}\cdot
\partial_{\boldsymbol{x}}\boldsymbol{e}_{i})\right]\cdot\partial_{\boldsymbol{x}}
+\frac{1}{2} \mathsf{g}^{-1}:\partial_{\boldsymbol{x}}\otimes\partial_{\boldsymbol{x}}\right\}f
\end{eqnarray}
( $\boldsymbol{b}\cdot\partial_{\boldsymbol{x}}\equiv \boldsymbol{b}^{\alpha_{1}}
\partial_{\boldsymbol{x}^{\alpha_{1}}}$ and $\mathsf{A}:\mathsf{B}\equiv 
\mathrm{Tr}\mathsf{A}^{\dagger}\mathsf{B}$). 
The last equality in (\ref{gen:Stratonovich}) motivates the identification of the inverse of the 
metric as the diffusion tensor of the process $\boldsymbol{\xi}$. %From the probabilistic point of view, 
A diffusion process is fully specified by the knowledge of the conditional expectation 
of its increments up to second order. %, which 
Under the present hypotheses these expectation values are
\begin{eqnarray}
\label{dp:drift}
\lefteqn{
\lim_{\dt \downarrow 0}\mathrm{E}_{\boldsymbol{x},t}\left\{\frac{\boldsymbol{\xi}_{t+\dt}
-\boldsymbol{\xi}_{t}}{\dt}\right\}=}
\nonumber\\&&
\left(\boldsymbol{b}_{S}+\frac{1}{2}(\boldsymbol{e}_{i}\cdot
\partial_{\boldsymbol{x}}\boldsymbol{e}_{i})\right)(\boldsymbol{x},t):=\boldsymbol{b}_{I}(\boldsymbol{x},t)
\end{eqnarray}
the ``It\^o drift'', and
\begin{eqnarray}
\label{dp:diffusion}
\lim_{\dt \downarrow 0}\mathrm{E}_{\boldsymbol{x},t}\left\{\frac{(\boldsymbol{\xi}_{t+\dt}-\boldsymbol{\xi}_{t})
\otimes(\boldsymbol{\xi}_{t+\dt}-\boldsymbol{\xi}_{t})}{\dt}\right\}
=\mathsf{g}^{-1}(\boldsymbol{x})
\end{eqnarray} 
the diffusion tensor. It is therefore physically justified to take (\ref{dp:drift}) and
(\ref{dp:diffusion}) as the \emph{data} specifying a stochastic differential equation.
We are now in the position to pinpoint two disadvantages related to the use of 
(\ref{sde:Stratonovich}). For any $\mathsf{O}\in O(d)$, the group of orthogonal matrices, 
the vector-valued Wiener processes $\boldsymbol{\beta}:=[\beta_{1},\dots,\beta_{d}]$ and $\mathsf{O}\boldsymbol{\beta}$ are statistically equivalent. 
Let us denote with $O(\mathbb{M})$ the collection of $d$-tuples 
$[\boldsymbol{x},\boldsymbol{e}_{1}(\boldsymbol{x}),\dots,
\boldsymbol{e}_{d}(\boldsymbol{x})]$ attached to any $\boldsymbol{x}\in\mathbb{M}$. 
In the language of differential geometry, this collection forms the ``bundle'' of 
orthonormal frames specified by the triple $(O(\mathbb{M}),O(d),\mathbb{M})$ 
\cite{IkMa79,IkedaWatanabe}. Since only (\ref{dp:diffusion}) is observable, any element
of the bundle must provide an equivalent description of the statistics of the process
$\boldsymbol{\xi}$.
%We see then that the above discussion pinpoints two 
%we can pinpoint two
%disadvantages of Stratonovich calculus in the form (\ref{sde:Stratonovich}). 
%To this goal, let us denote %, following the standard differential geometry notation 
%with $O(\mathbb{M})$ the set of $d$-tuples 
% $[\boldsymbol{x},\boldsymbol{e}_{1}(\boldsymbol{x}),\dots,
%\boldsymbol{e}_{d}(\boldsymbol{x})]$ attached to any $\boldsymbol{x}\in\mathbb{M}$. 
%We see then that the Stratonovich stochastic differential equation (\ref{sde:Stratonovich}) 
%depends upon the choice of the orthonormal
%frame in the ``bundle'' %of orthonormal frames 
%specified by the triple $(O(\mathbb{M}),O(d),\mathbb{M})$ \cite{IkMa79,IkedaWatanabe}. 
Equation (\ref{sde:Stratonovich}), however, is not invariant for different choices of 
orthonormal frames in the bundle. More explicitly, the same statistics 
can be equivalently described by the Stratonovich drift 
$\boldsymbol{b}_{S}$ or by its ``gauge'' transform
\begin{eqnarray}
\label{}
\boldsymbol{b}_{S}[\mathsf{O}]=\boldsymbol{b}_{S}-\frac{1}{2}
(\mathsf{O}_{i\,i_{1}}(\boldsymbol{e}_{i_{1}}\cdot\partial_{\boldsymbol{x}})
\mathsf{O}_{i\,i_{2}})\boldsymbol{e}_{i_{2}}
\end{eqnarray}
This fact was already noted long ago in connection to 
the inquiry of the covariant path-integral 
representation of stochastic processes and Euclidean Quantum Mechanics over curved 
manifolds (see e.g. \cite{It78,Gr85} and \cite{MG03} for a more complete list of references). 
A second disadvantage is that the natural, coordinate independent, definition of the
generator of $\boldsymbol{\xi}$ acting on \emph{scalar functions} is \cite{Ko37,Ito62}
\begin{eqnarray}
\label{gen:covariant}
\mathfrak{L}=\left(\boldsymbol{b}\cdot\partial_{\boldsymbol{x}}+\frac{1}{2}\Delta_{LB}\right)\,f
\end{eqnarray}
with $\Delta_{LB}$ the Laplace-Beltrami operator and $\boldsymbol{b}$ a vector field which we will 
refer to as the covariant drift. In general, it is not possible to identify
globally on $\mathbb{M}$ the Stratonovich drift with the covariant one i.e. $\boldsymbol{b}_{S}=\boldsymbol{b}$ \emph{unless} 
the integrability condition
\begin{eqnarray}
\label{str:integrability}
\boldsymbol{v}_{1}\cdot\partial_{\boldsymbol{x}}\langle\boldsymbol{v}_{2},\boldsymbol{e}\rangle_{\mathsf{g}}=
\boldsymbol{v}_{2}\cdot\partial_{\boldsymbol{x}}\langle\boldsymbol{v}_{1},\boldsymbol{e}\rangle_{\mathsf{g}}
\end{eqnarray}
is satisfied for any \emph{constant} vectors $\boldsymbol{v}_{1}$, $\boldsymbol{v}_{2}$. 
To interpret (\ref{str:integrability}) we observe that it is satisfied if there exists a 
collection of scalar functions $\left\{G_{i}\right\}_{i=1}^{d}$ such that
\begin{eqnarray}
\label{str:integrability2}
\partial_{\boldsymbol{x}}G_{i}=\mathsf{g}\cdot\boldsymbol{e}_{i}
\end{eqnarray}
whence it follows that 
\begin{eqnarray}
\label{}
\langle\boldsymbol{e}_{i},\dif\boldsymbol{\xi}_{t}\rangle_{\mathsf{g}}=\dif G_{i}(\boldsymbol{\xi}_{t})
\end{eqnarray}
meaning that there exists a change of variables turning in (\ref{gen:Stratonovich}) the frame fields
$\left\{\boldsymbol{e}_{i}\right\}_{i=1}^{d}$ into the canonical basis of $\mathbb{R}^{d}$. In other 
words, (\ref{str:integrability2}) states the existence of a priviledged choice of global coordinates 
for which the noise becomes additive. In general, however, (\ref{str:integrability}) is not 
satisfied since
\begin{eqnarray}
\label{}
\Delta_{LB}f=\mathsf{g}^{-1}:\partial_{\boldsymbol{x}}\otimes\partial_{\boldsymbol{x}}f-
(\mathsf{\Gamma}:\mathsf{g}^{-1})\cdot\partial_{\boldsymbol{x}}f
\end{eqnarray}
with $\mathsf{\Gamma}$ the Christoffel symbols of the Levi-Civita connection on $\mathbb{M}$. 
In \cite{Ito62} It\^o gave the expression in local coordinates of a stochastic differential 
equation (in It\^o sense) associated to (\ref{gen:covariant})
\begin{eqnarray}
\label{sde:Ito}
\dif\boldsymbol{\xi}_{t}=\left(\boldsymbol{b}-\frac{1}{2}\mathsf{\Gamma}:\mathsf{g}^{-1}\right)\dt
+\boldsymbol{e}_{i}\dif\beta_{i;t}
\end{eqnarray}
A straightforward calculation shows then that (\ref{sde:Stratonovich}) is equivalent to
(\ref{sde:Ito}) if (\ref{str:integrability}) is satisfied. The conclusion %of the above discussion 
%is that Stratonovich calculus alone is not sufficient to achieve a covariant 
%description of a non-equilibrium dynamics by means of stochastic differential equations.
is that the Stratonovich equation (\ref{sde:Stratonovich}) does not provide a 
coordinate-independent description of the dynamics.
Furthermore, inspection of (\ref{sde:Ito}) shows that %it is incorrect to identify
%the conditional expectation (\ref{dp:drift}), 
the ``It\^o drift'' cannot transform  as vector field under general change of coordinates.
%Fortunately, stochastic analysis provides us with a construction, 
The Eells-Elworthy-Malliavin development map 
(see e.g. \cite{IkedaWatanabe,Hsu02,Baudoin} for rigorous \emph{and} pedagogic derivations)
obviates these disadvantages.
The idea is that a Wiener process can be path-wise constructed on $\mathbb{M}$ by ``rolling'' 
the manifold along the realizations of an Euclidean Brownian motion. Mathematically this
means that the Wiener process should be constructed as the solution of the \emph{system}
of Stratonovich differential equations  
\begin{subequations}
\label{sde:devmap}
\begin{eqnarray}
\label{sde:devmap1}
\dif \boldsymbol{\omega}_{t}=\boldsymbol{e}_{i}\str{}\dif\beta_{i;t}
\end{eqnarray}
\begin{eqnarray}
\label{sde:devmap2}
\dif\boldsymbol{e}_{i;t}=-\mathsf{\Gamma}:\boldsymbol{e}_{i;t}\str{\otimes}\dif \boldsymbol{\omega}_{t}
\end{eqnarray}
\end{subequations}
As above, the $\left\{\beta_{i}\right\}_{i=1}^{d}$ are a collection of independent Wiener
processes. Were they constant vectors, (\ref{sde:devmap}) could be couched 
in the form of a geodesic equation. Furthermore, upon converting (\ref{sde:devmap1}) to the 
It\^o representation we recover (\ref{sde:Ito}) for vanishing covariant drift. An important 
further consequence of (\ref{sde:devmap1}) is that we can identify a stochastic process $\boldsymbol{\xi}$ 
as a local $\mathbb{M}$-valued martingale if in any local chart it satisfies 
\cite{Mey81,IkedaWatanabe,Nor92}
\begin{eqnarray}
\label{}
\lefteqn{
\dif \boldsymbol{\xi}_{t}+\frac{1}{2}\mathsf{\Gamma}:\dif \boldsymbol{\xi}_{t}\otimes
\dif \boldsymbol{\xi}_{t}\overset{law}{\equiv}
}
\nonumber\\&&
\hspace{-0.2cm}
\dif \boldsymbol{\xi}_{t}+\frac{1}{2}\mathsf{\Gamma}:\mathsf{g}^{-1}\dt=\mbox{local Euclidean martingale}
\end{eqnarray}
Three remarks are in order before concluding this short discussion of background results 
from stochastic analysis.
First, (\ref{sde:devmap}) admits a straightforward extension to semi-martingales, simply by introducing 
a drift term into (\ref{sde:devmap1}). We do not need to take this step here explicitly, 
since we can combine Girsanov formula and the development map (\ref{sde:devmap}) to take into account 
drift terms \cite{IkMa79}:
\begin{eqnarray}
\label{Girsanov:forward}
\lefteqn{
\frac{d\mathrm{P}_{\boldsymbol{\xi}}}{d\mathrm{P}_{\boldsymbol{\omega}}}(\boldsymbol{\omega})=
e^{\int_{\ti}^{\tf}\left\{\langle\boldsymbol{b}\,,\boldsymbol{e}_{i;t}\dif \beta_{i;t}\rangle_{\mathsf{g}}
-\frac{
\parallel\boldsymbol{b}\parallel_{\mathsf{g}}^{2}
%\langle\boldsymbol{b}\,,\boldsymbol{b}\rangle
\,\dt}{2}
\right\}(\boldsymbol{\omega}_{t},t)}
}
\nonumber\\&&
=e^{\int_{\ti}^{\tf}\left\{\langle \dif \boldsymbol{b}\str{\,,}\boldsymbol{\omega}_{t}\rangle_{\mathsf{g}}
-\frac{(\nabla\cdot\boldsymbol{b}+
\parallel\boldsymbol{b}\parallel_{\mathsf{g}}^{2}
%\langle\boldsymbol{b}\,,\boldsymbol{b}\rangle
)\,\dt}{2}
\right\}(\boldsymbol{\omega}_{t},t)}
\end{eqnarray}
for
\begin{eqnarray}
\label{}
\nabla\cdot\boldsymbol{b}=\frac{1}{\sqrt{ |g|}}\partial_{\boldsymbol{x}^{\alpha}}\,(\sqrt{ |g|}\,\boldsymbol{b}^{\alpha})
\end{eqnarray}
The symbol $\nabla$ betokens here and in what follows the covariant derivative operation
compatible with the metric $\mathsf{g}$. The stochastic integral in the first row of (\ref{Girsanov:forward})
is in It\^o sense. The use of the time-symmetric Stratonovich integral
in the second row of (\ref{Girsanov:forward}) exhibits that the argument of the exponential 
transforms indeed as a scalar under a change of coordinates.
Second, if the metric is time dependent we must include a drift term $1/2 \partial_{t}\,\mathsf{g}^{-1}$ in 
(\ref{sde:devmap2}) to preserve the metric compatibility of the connection. Finally \cite{Ko37}
the adjoint of  (\ref{gen:covariant})
\begin{eqnarray}
\label{}
\mathfrak{L}^{\dagger}f=-\nabla\cdot\boldsymbol{b}f+\frac{1}{2} \Delta_{LB}f
\end{eqnarray}
is defined with respect to the invariant Riemann volume:
\begin{eqnarray}
\label{}
\dif_{\mathbb{M}}^{d}x:=d^{d}x\,\sqrt{ \det\mathsf{g}}(\boldsymbol{x})
\end{eqnarray}
The covariant density $\mathrm{k}$ is then related to the transition probability density with 
respect to the Lebesgue measure $\mathrm{p}$ by the equation
\begin{eqnarray}
\label{}
\mathrm{p}(\boldsymbol{x}_{2},t_{2}|\boldsymbol{x}_{1},t_{1})
=\sqrt{ \det \mathsf{g}}(\boldsymbol{x}_{2})\,\mathrm{k}(\boldsymbol{x}_{2},t_{2}|\boldsymbol{x}_{1},t_{1})
\end{eqnarray}
for any coordinates $\boldsymbol{x}_{i}$, $i=1,2$ in a local chart $\mathbb{U}$, and any $t_{2}\,\geq\,t_{1}$ 
in $[\ti,\tf]$.

\section{Kullback-Leibler divergence and time-reversal}

Our aim is now to derive with the help of the development map the covariant expression of time 
reversal relations which can be used as bridge relations between the theory of stochastic 
processes and finite-time thermodynamics of small systems. Under rather general smoothness assumptions 
\cite{Nelson01,Nelson85}, if we know the probability $\mathrm{m}=\sqrt{ |\mathsf{g}|}\,\mathrm{n}$ and transition probability densities
 $\mathrm{p}=\sqrt{ |\mathsf{g}|}\,\mathrm{k}$. 
of a semi-martingale process $\boldsymbol{\xi}$ in the full time horizon $[\tf\,,\ti]$, we can construct the time 
reversed time evolution by requiring for any $\ti\,\leq\, t_{1}\,\leq\, t_{2}\,\leq\, \tf$
\begin{eqnarray}
\label{}
\mathrm{n}(\boldsymbol{x}_{2},t_{2})\,\mathrm{k}^{(r)}(\boldsymbol{x}_{1},t_{1}|\boldsymbol{x}_{2},t_{2})=
\mathrm{k}(\boldsymbol{x}_{2},t_{2}|\boldsymbol{x}_{1},t_{1})\,\mathrm{n}(\boldsymbol{x}_{1},t_{1})
\end{eqnarray}
If $\boldsymbol{\xi}$ is adapted to a $\mathbb{M}$-valued Wiener process (\ref{sde:devmap}), we can use Girsanov formula
(\ref{Girsanov:forward}) to evaluate averages of non-anticipative functionals
of $\boldsymbol{\xi}$ as averages with respect to $\boldsymbol{\omega}$. As noticed in \cite{Mey82}, we can also take advantage of
the invariance under time reversal of the Wiener process to express the density of a \emph{backward}
process $\boldsymbol{\tilde{\xi}}=\left\{\boldsymbol{\tilde{\xi}}_{t}\,;\,t\,\in\,[\ti\,,\tf]\right\}$:
\begin{eqnarray}
\label{Girsanov:backward}
\lefteqn{
\frac{\dif \mathrm{P}_{\boldsymbol{\tilde{\xi}}}}{\dif \mathrm{P}_{\boldsymbol{\omega}}}(\boldsymbol{\omega})=
e^{\int_{\ti}^{\tf}\left\{\langle\boldsymbol{\tilde{b}}\,,\boldsymbol{e}_{i;t}\oti{\,}\dif \beta_{i;t}\rangle_{\mathsf{g}}
-\frac{
%\langle\tilde{\boldsymbol{b}}\,,\tilde{\boldsymbol{b}}\rangle_{\mathsf{g}}
\parallel\boldsymbol{\tilde{b}}\parallel_{\mathsf{g}}^{2}
\dt}{2}
\right\}(\boldsymbol{\omega}_{t},t)}
}
\nonumber\\&&
=e^{\int_{\ti}^{\tf}\left\{\langle \dif \tilde{\boldsymbol{b}}\str{\,,}\boldsymbol{\omega}_{t}\rangle_{\mathsf{g}}
+\frac{(\nabla\cdot\boldsymbol{b}-
\parallel\boldsymbol{\tilde{b}}\parallel_{\mathsf{g}}^{2}
%\langle\tilde{\boldsymbol{b}}\,,\tilde{\boldsymbol{b}}\rangle_{\mathsf{g}}
)\dt}{2}
\right\}(\boldsymbol{\omega}_{t},t)}
\end{eqnarray}
The symbol $\oti{\,}$ appearing in the first row of (\ref{Girsanov:backward}) indicates that the stochastic integral 
is defined by taking the limit of Riemann sums sampled according to the 
post-point discretization; in other words, the first stochastic integral in (\ref{Girsanov:backward}) 
is an It\^o integral with respect to the filtration of the ``future'' while the second
is a Stratonovich integral. 
The requirement $\boldsymbol{\xi}\overset{law}{=}\boldsymbol{\tilde{\xi}}$ translates into
\begin{eqnarray}
\label{Girsanov:inversion}
\mathrm{n}(\boldsymbol{\omega}_{t_{2}},t_{2})
\frac{\dif \mathrm{P}_{\boldsymbol{\tilde{\xi}}}}{\dif \mathrm{P}_{\boldsymbol{\omega}}}(\boldsymbol{\omega})=
\mathrm{n}(\boldsymbol{\omega}_{t_{1}},t_{1})
\frac{\dif \mathrm{P}_{\boldsymbol{\xi}}}{\dif \mathrm{P}_{\boldsymbol{\omega}}}(\boldsymbol{\omega})
\end{eqnarray}
whence we immediately recover the classical result
\begin{eqnarray}
\label{drift:backwards}
\boldsymbol{b}(\boldsymbol{x},t)=\tilde{\boldsymbol{b}}(\boldsymbol{x},t)
+\mathsf{g}^{-1}\cdot\partial_{\boldsymbol{x}}\mathrm{n}(\boldsymbol{x},t)
\end{eqnarray}
Both sides of the equation are now well-defined vector fields. We thus see that the \emph{same}
probability measure admits a forward and backward dynamics representation in terms
of two \emph{different} drift vector fields. A natural way to compare the two vector fields
with respect to the same filtration is to introduce  a \emph{forward, auxiliary} process 
$\bar{\boldsymbol{\xi}}$  \emph{absolutely continuous} with respect to $\boldsymbol{\xi}$
defined by the replacement 
\begin{eqnarray}
\label{drift:auxiliary}
\boldsymbol{b}\mapsto \boldsymbol{\bar{b}}:=\,-\,\boldsymbol{\tilde{b}}
\end{eqnarray}
in the stochastic differential equation driving $\boldsymbol{\xi}$. The rationale behind
(\ref{drift:auxiliary}) is to combine the stochastic with deterministic time 
reversal. As a consequence, for a gradient--type drift (see equation (\ref{td:grad}) 
below) at equilibrium (\ref{drift:auxiliary}) reduces to the identity $\boldsymbol{b}=\boldsymbol{\bar{b}}$.
In general a natural quantifier of the ``drift discrepancy'' 
is obtained by considering the Kullback-Leibler divergence \cite{KuLe51}
of the process $\bar{\boldsymbol{\xi}}$ with respect to $\boldsymbol{\xi}$:
\begin{eqnarray}
\label{KuLe:def}
\lefteqn{
\mathcal{K}(\mathrm{P}_{\boldsymbol{\bar{\xi}}}||\mathrm{P}_{\boldsymbol{\xi}}):=\mathrm{E}^{(\boldsymbol{\xi})}
\ln\frac{\dif \mathrm{P}_{\boldsymbol{\xi}}}{\dif \mathrm{P}_{\boldsymbol{\bar{\xi}}}}(\boldsymbol{\xi})
}
\nonumber\\&&
=\mathrm{E}^{(\boldsymbol{\omega})}\frac{\dif \mathrm{P}_{\boldsymbol{\xi}}}{\dif \mathrm{P}_{\boldsymbol{\omega}}}(\boldsymbol{\omega})
\ln\left(\frac{\dif \mathrm{P}_{\boldsymbol{\xi}}}{\dif \mathrm{P}_{\boldsymbol{\omega}}}
\frac{\dif \mathrm{P}_{\boldsymbol{\omega}}}{\dif \mathrm{P}_{\boldsymbol{\bar{\xi}}}}\right)(\boldsymbol{\omega})
\end{eqnarray}
The notation $\mathrm{E}^{(\boldsymbol{\xi})}$ ( $\mathrm{E}^{(\boldsymbol{\omega})}$ ) emphasizes that the average is 
with respect to the measure of
$\boldsymbol{\xi}$ ($\boldsymbol{\omega}$). The Kullback-Leibler divergence is by construction a positive definite quantity. 
Direct evaluation of (\ref{KuLe:def}) yields
\begin{eqnarray}
\label{KuLe:explicit}
\lefteqn{
\mathcal{K}(\mathrm{P}_{\boldsymbol{\bar{\xi}}}||\mathrm{P}_{\boldsymbol{\xi}})
=\mathrm{E}^{(\boldsymbol{\xi})}\int_{\ti}^{\tf}\langle\boldsymbol{b}
+\boldsymbol{\tilde{b}}\,\str{,}\,\dif\boldsymbol{\xi}_{t}\rangle_{\mathsf{g}}
}
\nonumber\\&&
-\mathrm{E}^{(\boldsymbol{\xi})}\int_{\ti}^{\tf}\dt\,\frac{\nabla_{\boldsymbol{\xi}_{t}}\cdot(\boldsymbol{b}+\boldsymbol{\tilde{b}})
+\parallel\boldsymbol{b}\parallel_{\mathsf{g}}^{2}-\parallel\boldsymbol{\tilde{b}}\parallel_{\mathsf{g}}^{2}}{2}
%-\mathrm{E}^{(\boldsymbol{\xi})}\int_{\ti}^{\tf}\dt\left\{\nabla\cdot\boldsymbol{b}
%+\langle\boldsymbol{b}\,,\mathsf{g}^{-1}\cdot\partial_{\boldsymbol{\xi}_{t}}\ln\mathrm{n}\rangle_{\mathsf{g}}
%-\frac{\nabla\cdot\partial_{\boldsymbol{\xi}_{t}}\ln\mathrm{n}
%+\langle\mathsf{g}^{-1}\cdot\partial_{\boldsymbol{\xi}_{t}}\ln\mathrm{n}\,,
%\mathsf{g}^{-1}\cdot\partial_{\boldsymbol{\xi}_{t}}\ln\mathrm{n}\rangle_{\mathsf{g}}}{2}\right\}
\end{eqnarray}
Upon inserting (\ref{drift:backwards}) into (\ref{KuLe:explicit}) and using
probability conservation and the covariant Fokker-Planck equation
\begin{eqnarray}
\label{FP}
(\partial_{t}-\mathfrak{L}^{\dagger})\mathrm{n}=0
\end{eqnarray}
we can prove that the Riemann integral in (\ref{KuLe:explicit}) vanishes. 
The integrand of the Stratonovich stochastic integral is instead proportional
to the \emph{current velocity} \cite{Nelson01,Nelson85} of the process $\boldsymbol{\xi}$:
\begin{eqnarray}
\label{}
\boldsymbol{v}=\frac{\boldsymbol{b}+\boldsymbol{\tilde{b}}}{2}
\end{eqnarray}
The current velocity enjoys two important properties. First, it plays for expectation 
values of Stratonovich line integrals the same role as the It\^o drift (\ref{dp:drift}) 
for expectation values of It\^o line integrals: for any smooth, mean square integrable 
vector field $\boldsymbol{h}$ the equality
\begin{eqnarray}
\label{cv:expectation}
\mathrm{E}^{(\boldsymbol{\xi})}\int_{\ti}^{\tf}\langle\boldsymbol{h}\,\str{,}\,\dif\boldsymbol{\xi}\rangle_{\mathsf{g}}=
\mathrm{E}^{(\boldsymbol{\xi})}\int_{\ti}^{\tf}\dt\,\langle\boldsymbol{h}\,,\boldsymbol{v}\rangle_{\mathsf{g}}
\end{eqnarray}
holds true under the assumption that integrations by parts do not bring about
boundary terms.
Second, in terms of the current velocity the covariant Fokker-Planck equation reduces
to deterministic mass conservation equation:
\begin{eqnarray}
\label{cv:mt}
\partial_{t}\mathrm{n}+\nabla\cdot\boldsymbol{v}\,\mathrm{n}=0
\end{eqnarray}
Combining (\ref{cv:expectation}) with (\ref{drift:backwards}) and (\ref{FP}) 
we finally arrive at
\begin{eqnarray}
\label{KuLe:ep}
\mathcal{K}(\mathrm{P}_{\boldsymbol{\bar{\xi}}}||\mathrm{P}_{\boldsymbol{\xi}})
=2\,
\mathrm{E}^{(\boldsymbol{\xi})}\int_{\ti}^{\tf}\dt\,\parallel \boldsymbol{v}\parallel^{2}_{\mathsf{g}}
\end{eqnarray}
The right hand side of (\ref{KuLe:ep}) is proportional to the ``kinetic energy''
specified by the current velocity of the system. Its physical interpretation   
\cite{MaNeWy08,AuGaMeMoMG12} (see also \cite{Gar05,Sei05,ChGa07}) is that of 
\emph{entropy production} during the transformation. The way we arrived to 
(\ref{KuLe:ep}) differs to some extent from the one of \cite{Kur98,LeSp99,MaReMo00}. 
There the role of the Kullback-Leibler divergence is played by an ``action functional'' 
defined by contrasting the forward dynamics with its path-space time reversed defined 
by inverting the arrow of time $t\mapsto \tf+\ti-t$ in analogy to what was done to prove 
fluctuation theorems in hyperbolic dynamics \cite{GaCo95}. 
%Both the present and approach of \cite{Kur98,LeSp99,MaReMo00} can be generalized 
%to combine time-reveral with involutions of coordinate variables as in \cite{ChGa07} 
%the analysis whereof is, however, based on the stochastic dynamics described by 
%(\ref{sde:Stratonovich}). 
The entropy production relation can also be regarded as the continuum limit
under diffusive scaling of the analogous relation found in \cite{MGMePe12} for
Markov jump processes. 

\section{Stochastic thermodynamics}

Let us now examine the consequences of (\ref{KuLe:ep}) for optimal control in
thermodynamic functionals. We thus consider a dynamical system driven by 
time-dependent gradient-like drift
\begin{eqnarray}
\label{td:grad}
\boldsymbol{b}(\boldsymbol{x},t)=
-\mathsf{g}^{-1}(\boldsymbol{x})\cdot\partial_{\boldsymbol{x}}U(\boldsymbol{x},t)
\end{eqnarray}
which is a stylized model of a mechanical potential subject to an external 
control. The current velocity becomes in such a case
\begin{eqnarray}
\label{td:cv}
\boldsymbol{v}(\boldsymbol{x},t)=
-\mathsf{g}^{-1}(\boldsymbol{x})\cdot\partial_{\boldsymbol{x}}\left(U+\frac{1}{2}\ln\mathrm{n}\right)
(\boldsymbol{x},t)
\end{eqnarray}
Drawing from \cite{Se98}, we define the work done on the system as
\begin{eqnarray}
\label{td:work}
\lefteqn{
\mathcal{W}_{\tf,\ti}
=\mathrm{E}^{(\boldsymbol{\xi})}\int_{\ti}^{\tf}\dt\,\partial_{t}U(\boldsymbol{\xi}_{t},t)
=\mathrm{E}^{(\boldsymbol{\xi})}\left.U(\boldsymbol{\xi}_{t},t)\right|_{\ti}^{\tf}
}
\nonumber\\&&
-\mathrm{E}^{(\boldsymbol{\xi})}\int_{\ti}^{\tf} \langle d\boldsymbol{\xi}_{t}\,\str{,}\,
\mathsf{g}^{-1}(\boldsymbol{\xi}_{t})
\cdot\partial_{\boldsymbol{\xi}_{t}}
U(\boldsymbol{\xi}_{t},t)\rangle_{\mathsf{g}}
\hspace{2.0cm}
\end{eqnarray}
If we require the first law of thermodynamics to hold true, we must identify the 
heat as
\begin{eqnarray}
\label{td:heat}
\mathcal{Q}_{\tf,\ti}=
-\mathrm{E}^{(\boldsymbol{\xi})}\int_{\ti}^{\tf} 
\langle d\boldsymbol{\xi}_{t}\,\str{,}\,\mathsf{g}^{-1}(\boldsymbol{\xi}_{t})
\cdot\partial_{\boldsymbol{\xi}_{t}}
U(\boldsymbol{\xi}_{t},t)\rangle_{\mathsf{g}}
\end{eqnarray}
Under the present conventions, the local equilibrium condition is
\begin{eqnarray}
\label{}
\nabla_{\boldsymbol{x}}\cdot\mathsf{g}^{-1}\cdot\left(\mathrm{n}\,\partial_{\boldsymbol{x}}U
+\frac{1}{2}\partial_{\boldsymbol{x}}\mathrm{n}\right)=0
\end{eqnarray}
Correspondingly, we define the osmotic velocity \cite{Nelson01,Nelson85} as
\begin{eqnarray}
\label{td:ov}
\lefteqn{
\boldsymbol{u}(\boldsymbol{x},t):=
\frac{1}{2}(\boldsymbol{b}-\boldsymbol{\tilde{b}})(\boldsymbol{x},t)
}
\nonumber\\&&
=\frac{1}{2}\mathsf{g}^{-1}(\boldsymbol{x})
\cdot\partial_{\boldsymbol{x}}
\ln n(\boldsymbol{x},t)
\end{eqnarray}
and the scalar (i.e. coordinate independent) expression of the Gibbs-Shannon entropy
\begin{eqnarray}
\label{td:Shannon}
\lefteqn{
\mathcal{S}(t)=-\frac{1}{2}\mathrm{E}^{(\boldsymbol{\xi})}\ln\mathrm{n}(\boldsymbol{\xi}_{t},t)
}
\nonumber\\&&
\equiv\,-\,\frac{1}{2}\int_{\mathbb{M}}\dif_{\mathbb{M}}^dx\,(\mathrm{n}\,\ln
\mathrm{n})(\boldsymbol{x},t)
\end{eqnarray}
Note that the definition (\ref{td:Shannon}) of the Gibbs-Shannon entropy 
differs from the one given in \cite{AuMeMG12,AuGaMeMoMG12} 
by a prefactor $1/2$ dictated by the convention adopted here for the diffusion tensor,
and by an addend proportional to $\ln\sqrt{ |\mathsf{g}|}$ which does not transform as a scalar under 
change of coordinates.
If we now add and subtract the osmotic velocity to the integrand 
in (\ref{td:work}) and (\ref{td:heat}), use probability conservation and 
the identity (\ref{cv:expectation}), we arrive at the manifestly coordinate-independent
representations of the released heat
\begin{eqnarray}
\label{td:heat2}
\mathcal{Q}_{\tf,\ti}=-[\mathcal{S}(\tf)-\mathcal{S}(\ti)]
+\mathrm{E}^{(\boldsymbol{\xi})}\int_{\ti}^{\tf}\dt\,\parallel\boldsymbol{v}\parallel_{\mathsf{g}}^{2}
\end{eqnarray}
and of thermodynamic work
\begin{eqnarray}
\label{td:work2}
\mathcal{W}_{\tf,\ti}=\mathcal{F}(\tf)-\mathcal{F}(\ti)
+\mathrm{E}^{(\boldsymbol{\xi})}\int_{\ti}^{\tf}\dt\,\parallel\boldsymbol{v}\parallel_{\mathsf{g}}^{2}
\end{eqnarray}
In (\ref{td:work2}) the Helmholtz free energy of the system is
\begin{eqnarray}
\label{td:Helmoltz}
\lefteqn{
\mathcal{F}(t)\equiv\mathcal{U}(t)-\mathcal{S}(t)=
}
\nonumber\\&&
\mathrm{E}^{(\boldsymbol{\xi})}\left(U+\frac{1}{2}\ln \mathrm{n}\right)(\boldsymbol{\xi}_{t},t):=
-\,\mathrm{E}^{(\boldsymbol{\xi})}\psi(\boldsymbol{\xi}_{t},t)
\end{eqnarray}
The advantage to introduce here the \emph{scalar} quantity $\psi$ to denote the
free energy density emerges when contrasting the definitions of the current and 
osmotic velocities with (\ref{td:grad}). We recognize then that the current velocity
is the gradient of the Helmholtz free energy density: 
\begin{eqnarray}
\label{td:vpot}
\boldsymbol{v}(\boldsymbol{x},t)=
\mathsf{g}^{-1}(\boldsymbol{x})\cdot\partial_{\boldsymbol{x}}\psi(\boldsymbol{x},t)
\end{eqnarray} 
It is worth noticing that the Helmholtz free energy $\psi$ density  
(\ref{td:Helmoltz}) coincides with the potential similarly denoted in \cite{AuMeMG12} owing to a
cancellation of terms proportional to $\ln \sqrt{|\mathsf{g}|}$ between the individually non-coordinate-independent 
expressions of the internal energy and the Gibbs-Shannon entropy thereby used.  
Similar considerations, exploiting the time-independence of the Riemann metric also guarantee
that the expression for the released heat of \cite{AuMeMG12} does coincide with the one given 
here. The same is true for the work whose expression (\ref{td:work2}) has been obtained
also in \cite{BeGaJoLa12} using, however, arguments valid only in the weak noise limit.  

\section{Optimal control}

The expressions (\ref{td:heat2}) and (\ref{td:work2}) evince that the current velocity
formulation of Langevin thermodynamics is natural from the point of view of control theory.
There are at least two kinds of considerations supporting this claim. First, by (\ref{cv:mt})
the current velocity maps the original stochastic control problem into a deterministic 
control one. Such a mapping exists for any smooth diffusion but it is of limited practical
use if the drift is known a-priori and the problem is to derive the evolution of the density.
It becomes useful for control purposes when the drift is not known but must be determined 
by minimizing a ``cost'' functional over a suitable space of admissible controls. 
The second consideration is that the functional dependence of the entropy production on the current
velocity readily enforces the \emph{coercivity condition} (see e.g. \cite{FlemingSoner} pag. 33: 
convexity with growth faster than linear) which plays an important role in variational 
calculus to prove the existence of extremal solutions. 
A direct consequence of these considerations is the recovery of Jarzynski's  
\begin{eqnarray}
\label{td:Jarzynski}
\mathcal{W}_{\tf,\ti}\,
\geq\, \mathcal{F}(\tf)-\mathcal{F}(\ti)
\end{eqnarray}
and ``finite--time Landauer's'' \cite{Lan61} inequalities
\begin{eqnarray}
\label{td:Landauer}
\mathcal{Q}_{\tf,\ti}\,\geq\, -[\mathcal{S}(\tf)-\mathcal{S}(\ti)]
\end{eqnarray}
The interpretation of (\ref{td:Landauer}) as Landauer's inequality stems from 
the identification of the heat release with the entropy variation of the 
environment.
By (\ref{td:cv}) the current velocity vanishes at equilibrium. As a consequence, the
inequalities (\ref{td:Jarzynski}), (\ref{td:Landauer}) become tight for transitions 
described by a \emph{jump} between two ``equilibrium'' states. Physically, jumps are
mathematical mock-up's for transitions occurring at the fastest admissible time-scale.
As such, they are not suited to describe macroscopic control of a nano-system. 
It is therefore relevant to look for minima of thermodynamics indicators by restricting 
the space of admissible controls to those guaranteeing a smooth behavior of the
current velocity.
In order to achieve this goal let us recall that given a forward Markov process $\boldsymbol{\eta}$,
deterministic or stochastic, with generator $\mathfrak{G}_{[\boldsymbol{u}]}$ 
depending on a control $\boldsymbol{u}$, the canonical form of cost functionals considered in 
control theory \cite{FlemingSoner} is
\begin{eqnarray}
\label{control:cost}
\mathcal{A}(\boldsymbol{x},t)=\mathrm{E}_{\boldsymbol{x},t}^{(\boldsymbol{\eta})}\Psi(\boldsymbol{\eta}_{\tf})
+\mathrm{E}_{\boldsymbol{x},t}^{(\boldsymbol{\eta})}\int_{t}^{\tf}\dt\,L(\boldsymbol{\eta}_{t},t;\boldsymbol{u})
\end{eqnarray}
Standing some regularity assumptions, it is a-priori tenable to expect 
the ``cost functional'' $\mathcal{A}$ to admit a minimum over the space of smooth controls 
$\boldsymbol{u}$ if the ``running cost'' $L$ depends coercively on $\boldsymbol{u}$ and
the ``terminal cost'' $\Psi$ is some assigned function independent of $\boldsymbol{u}$.
The minimum 
\begin{eqnarray}
\label{}
J_{\star}(\boldsymbol{x},t)=\min_{\boldsymbol{u}}\mathcal{A}(\boldsymbol{x},t)
\end{eqnarray}
is usually referred to as the ``value function''.
Under these hypotheses, the solution of the Hamilton-Jacobi-Bellman equation
\begin{subequations}
\label{control:HJB}
\begin{eqnarray}
\label{control:HJB1}
\partial_{t}J_{\star}+\min_{\boldsymbol{u}}\left\{\mathfrak{G}_{[\boldsymbol{u}]}J_{\star}+L\right\}=0
\end{eqnarray}
\begin{eqnarray}
\label{control:HJB2}
J_{\star}(\boldsymbol{x},\tf)=\Psi(\boldsymbol{x})
\end{eqnarray}
\end{subequations}
corresponding to the smallest value of $\mathcal{A}$ (in case of multiple minima in
(\ref{control:HJB1}))  specifies the value of the optimal 
control $\boldsymbol{u}_{\star}$ for any $t$ in a \emph{closed} horizon $[\ti\,,\tf]$ if the resulting 
$\mathfrak{G}_{[\boldsymbol{u}_{\star}]}$ is a well-defined generator of a Markov process. 
The properties that a solution of (\ref{control:HJB}) must enjoy in order to satisfy 
such self-consistency condition are determined by so called \emph{verification theorems}
\cite{FlemingSoner,vHa07}.
Linearity of the Kolmogorov pair of equations governing a Markovian dynamics extends 
(\ref{control:HJB}) to expectations of $\mathcal{A}$ with respect to the measure of $\boldsymbol{\eta}$ evolving 
from non-localized initial density assigned at time $t=\ti$ \cite{GuMo83}. Namely, if we 
denote variations with respect to the control $\boldsymbol{u}$ by a ``prime symbol'', we can couch the 
variation of $\mathrm{E}^{(\boldsymbol{\eta}_{t})}\mathcal{A}$ into the form
\begin{eqnarray}
\label{}
\lefteqn{
(\mathrm{E}^{(\boldsymbol{\eta})}\mathcal{A})^{\prime}(\boldsymbol{\eta}_{\ti},\ti)=\int_{\mathbb{M}}
\dif_{\mathbb{M}}^{d}x\,(\mathrm{n}_{\boldsymbol{\eta}}^{\prime}\,\Psi)(\boldsymbol{x},t)
}
\nonumber\\&&
+\int_{\ti}^{\tf}\dt\int_{\mathbb{M}}\dif_{\mathbb{M}}^{d}x\,
\left(\mathrm{n}_{\boldsymbol{\eta}}^{\prime}\,L+\mathrm{n}_{\boldsymbol{\eta}}
L^{\prime}\right)(\boldsymbol{x},t)
\end{eqnarray}
If we now require the ``dynamic programming'' (non-homogeneous backward Kolmogorov) equation 
to hold for any admissible control
\begin{eqnarray}
\label{control:dp}
(\partial_{t}+\mathfrak{G}_{[\boldsymbol{u}]})\,J+L=0
\end{eqnarray}
we obtain for $\mathrm{n}_{\boldsymbol{\eta}}^{\prime}(\cdot,\ti)=0$
\begin{eqnarray}
\label{control:v1}
\lefteqn{
(\mathrm{E}^{(\boldsymbol{\eta})}\mathcal{A})^{\prime}=\int_{\mathbb{M}}
\dif_{\mathbb{M}}^{d}x\,[\mathrm{n}_{\boldsymbol{\eta}}^{\prime}\,(\Psi-J)](\boldsymbol{x},t)
}
\nonumber\\&&
+\int_{\ti}^{\tf}\dt\int_{\mathbb{M}}\dif_{\mathbb{M}}^{d}x\,\left\{
J\,[(\partial_{t}-\mathfrak{G}_{[\boldsymbol{u}]}^{\dagger})\mathrm{n}_{\boldsymbol{\eta}}]^{\prime}
\right\}(\boldsymbol{x},t)
\nonumber\\&&
+\int_{\ti}^{\tf}\dt\int_{\mathbb{M}}\dif_{\mathbb{M}}^{d}x\,\left\{
\mathrm{n}_{\boldsymbol{\eta}}(\mathfrak{G}_{[\boldsymbol{u}]}^{\prime}J+L^{\prime})\right\}(\boldsymbol{x},t)
\end{eqnarray}
The variation yields a stationary point if (\ref{control:HJB2}) is satisfied, 
the probability density evolves for any $\boldsymbol{u}$ according to the dynamics 
specified by the adjoint action of $\mathfrak{G}_{[\boldsymbol{u}]}$,
and the optimal control is fixed by the same condition as in (\ref{control:HJB1}). If, furthermore,
the generator is linear in the control $\boldsymbol{u}$, the convexity of $L$ readily implies
that the stationary point is a minimum
\begin{eqnarray}
\label{}
(\mathrm{E}^{(\boldsymbol{\eta})}\mathcal{A})^{\prime\prime}
=\int_{\ti}^{\tf}\dt\int_{\mathbb{M}}\dif_{\mathbb{M}}^{d}x\,\left(
\mathrm{n}_{\boldsymbol{\eta}}\,L^{\prime\prime}\right)(\boldsymbol{x},t)\geq 0
\end{eqnarray}
The dynamic programming equation (\ref{control:dp}) has the hydrodynamic interpretation
of a material derivative along the realizations of the Markov process $\boldsymbol{\eta}$.
Requiring it to hold a priori substantiate Bellman's idea that optimal control stems from a 
condition imposed \emph{locally} during the time evolution \cite{FlemingSoner,vHa07}. Within 
this framework, probability evolution must be regarded as the adjoint transport equation
which may-be non-local owing to the boundary conditions and must hold in the same time horizon
$[\ti,\tf]$ where (\ref{control:dp}) is defined. 
In \cite{GuMo83} it was also shown that the above chain of steps holds also if the running cost
$L$ depends upon derivatives of the control. This fact was originally exploited in \cite{AuMeMG11}
to unveil the relation between optimal thermodynamic control and hydrodynamic mass transport.
The formulation in terms of the current velocity renders, however, this extension no further needed
for thermodynamic control.

In the application of the control-theory toolkit to expressions such as (\ref{td:work2})
we are confronted with a subtle difficulty. While the entropy production specifies a well-defined 
coercive running cost, the interpretation of the terminal cost is more problematic. 
Namely, a candidate smooth optimal control must satisfy the stationarity condition
\begin{eqnarray}
\label{oc:sc}
\partial_{\boldsymbol{x}}J_{\star}+2\,\mathsf{g}\cdot\boldsymbol{v}=\partial_{\boldsymbol{x}}(J_{\star}+2\,\psi)=0
\end{eqnarray}
and the terminal condition (we suppose that no optimization is carried out over 
the initial state)
\begin{eqnarray}
\label{oc:tc}
J_{\star}(\boldsymbol{x},\tf)=\Psi(\boldsymbol{x})
\end{eqnarray}
Contrasting (\ref{oc:sc}) and (\ref{oc:tc}) with (\ref{td:vpot}) it is evident that
the terminal condition \emph{cannot} be interpreted as the end-horizon value of the free 
energy density of a \emph{smooth optimal} protocol since in general 
$\psi(\cdot,\tf)\neq \Psi(\cdot)$. Furthermore, a process jumping at
$\tf^{-}$ from a state of \emph{non-vanishing} current velocity \emph{cannot} 
be considered \emph{optimal}, independently of the prescription adopted for the probability 
density at time $\tf$. Namely, if we enlarge the space
of admissible protocols to encompass jump processes, we know a-priori that the infimum
is attained by an instantaneous transition between ``equilibria''. In this latter
case the current velocity and the running cost vanish identically in the full control
horizon and the work coincides with the free energy difference. The conclusion is that
the ``optimal work'' protocol proposed in \cite{ScSe07} \emph{cannot} be justified 
using Hamilton-Jacobi-Bellman theory and Langevin dynamics alone. It appears instead to 
describe an optimal control strategy if the Langevin dynamics is embedded 
into an higher order Markovian dynamics \cite{AuMeMG12} following ideas closely
reminiscent of the ``valley method'' (see e.g. \cite{AoKiOkSaWa99}). The ``valley method'' 
is a technique which aims at justifying the stationary point approximation to path integrals 
when no exact classical field configuration can match the required boundary conditions. 
It is worth stating clearly that if the terminal cost (\ref{oc:tc}) is interpreted only as
the variation of an external potential without explicit relation with $\psi$ in the control horizon
$[\ti\,,\tf]$ then the protocol found in \cite{ScSe07,AuMeMG12} is optimal according to 
standard verification theorems (see e.g. discussion in sections III.5 to III.8 of \cite{FlemingSoner}).

A control problem which lends itself to a more transparent physical interpretation
is that of the minimization of the entropy production between assigned probability densities 
at the end of the control horizon. The analytical and numerical treatment of this
 problem have been inquired in details in \cite{AuGaMeMoMG12}. It is worth here
to draw the attention some aspects of this problem not discussed in \cite{AuMeMG12,AuGaMeMoMG12}.  
By (\ref{KuLe:ep}) the problem can be equivalently formulated in terms of the 
minimization of the Kullback-Leibler divergence associated to a time reversal operation.
It is instructive to contrast thermodynamic entropy production minimization with the control problem
defining the so-called  ``Schr\"odinger diffusion'' see e.g. \cite{Sc31,Da91,FiHo04,MiTh06}. 
Given two probability densities at the end of a control horizon 
and a reference diffusion process, the Schr\"odinger diffusion problem determines 
the smoothly interpolating diffusion obtained from the reference process by a deformation 
of the drift under the requirement that the Kullback-Leibler divergence between the 
two processes be at a minimum value. 
If the reference process is the Wiener process then the Schr\"odinger diffusion corresponds to treat as 
running cost instead of the entropy production an analogous quantity in which the current 
velocity is replaced by the forward drift of the process.
Correspondingly, the associated control problem is turned from deterministic to stochastic.
A widely used approach to deterministic control is to prove existence and uniqueness 
of solutions in viscosity sense \cite{FlemingSoner}. In simplest cases this means 
constructing solutions as the inviscid limit of an ultraviolet regularization of the
Hamilton-Jacobi equation by adding a Laplacian. The Monge-Amp\`ere-Kantorovich method \cite{MoSo07}
applied in \cite{AuGaMeMoMG12} to study nucleation at minimum entropy production is an example
of this general ideology. This observation allows us to attribute a direct physical 
interpretation also to the regularized entropy production minimization problem. 
Namely, in the presence of any finite viscosity 
we can interpret the minimizer of the Kullback--Leibler divergence (\ref{KuLe:ep}) as the drift solving 
an associated Schr\"odinger diffusion problem.

\section{Conclusions}

In conclusion, we showed how a coordinate independent formalism for stochastic differential equations
provides a convenient formulation of control problems arising in stochastic thermodynamics. In doing so
we restricted the attention to time--independent diffusion tensors. This is not too restrictive 
under the hypothesis that mechanical forces and control parameters are most naturally encapsulated 
in the drift field whilst the diffusion coefficient contains purely geometric information. 
We also analyzed how the local nature of Bellman principle affects the optimal
control equations. Preserving the adjoint structure of the Kolmogorov pair requires for example 
that jumps in the protocol which is governed by the backward Kolmogorov equation bring about jumps in the forward 
Kolmogorov equation governing the probability density evolution. Hence, it is an essential modeling 
question to assess a-priori which is the physically relevant space of admissible protocols and whether
and how to describe the procedures to switch on and off the optimal protocol at the end of the control horizon.

\section{Acknowledgments}

It is a pleasure to thank Carlos Mej\'ia--Monasterio for discussions and useful comments on this manuscript.
The work of PMG is supported by by  the  Center  of
Excellence  ``Analysis and  Dynamics''of the  Academy of  Finland.

\addcontentsline{toc}{section}{Bibliography}
\bibliography{/home/paolo/RESEARCH/BIBTEX/jabref}{} 

\begin{thebibliography}{10}

\bibitem{AoKiOkSaWa99}
H.~Aoyama, H.~Kikuchi, I.~Okouchi, M.~Sato, and S.~Wada.
\newblock {Valley Views: Instantons, Large Order Behaviors, and Supersymmetry}.
\newblock {\em Nuclear Physics B}, 553(3):644--710, August 1999,
  arXiv:hep-th/9808034.

\bibitem{AuGaMeMoMG12}
E.~Aurell, K.~Gaw\c{e}dzki, C.~Mej\'ia-Monasterio, R.~Mohayaee, and
  P.~Muratore-Ginanneschi.
\newblock {Refined Second Law of Thermodynamics for fast random processes}.
\newblock {\em Journal of Statistical Physics}, 147(3):487--505, April 2012,
  arXiv:1201.3207.

\bibitem{AuMeMG11}
E.~Aurell, C.~Mej\'ia-Monasterio, and P.~Muratore-Ginanneschi.
\newblock {Optimal protocols and optimal transport in stochastic
  thermodynamics}.
\newblock {\em Physical Review Letters}, 106(25):250601, June 2011,
  arXiv:1012.2037.

\bibitem{AuMeMG12}
E.~Aurell, C.~Mej\'ia-Monasterio, and P.~Muratore-Ginanneschi.
\newblock {Boundary layers in stochastic thermodynamics}.
\newblock {\em Physical Review E}, 85(2):020103(R), Februray 2012,
  arXiv:1111.2876.

\bibitem{Baudoin}
F.~Baudoin.
\newblock {\em An introduction to the geometry of stochastic flows}.
\newblock Imperial College Press, 2004.

\bibitem{BeGaJoLa12}
L.~Bertini, D.~Gabrielli, G.~Jona-Lasinio, and C.~Landim.
\newblock {Thermodynamic transformations of nonequilibrium states}.
\newblock {\em arXiv preprint}, 2012,  arXiv:1206.2412.

\bibitem{ChGa07}
R.~Ch{\'e}trite and K.~Gaw\c{e}dzki.
\newblock Fluctuation relations for diffusion processes.
\newblock {\em Communications in Mathematical Physics}, 282:469--518, 2007,
  arXiv:0707.2725.

\bibitem{Cou11}
K.~A. Coulibaly-Pasquier.
\newblock {Brownian motion with respect to time-changing Riemannian metrics,
  applications to Ricci flow}.
\newblock {\em Annales de l'Institut Henri Poincar\'e, Probabilit\'es et
  Statistiques}, 47(2):515--538, 2011,
  oai:hal.archives-ouvertes.fr:hal-00352805.

\bibitem{Da91}
P.~Dai~Pra.
\newblock {A stochastic control approach to reciprocal diffusion processes}.
\newblock {\em Applied Mathematics and Optimization}, 23(1):313--329, 1991.

\bibitem{FiHo04}
R.~Filliger and M.-O. Hongler.
\newblock {Relative Entropy and Efficiency Measure for diffusion-mediated
  transport processes}.
\newblock {\em Journal of Physics A}, 38:1247--1255, 2004.

\bibitem{FlemingSoner}
W.~H. Fleming and M.~H. Soner.
\newblock {\em Controlled {Markov} processes and viscosity solutions},
  volume~25 of {\em Stochastic modelling and applied probability}.
\newblock Springer, 2nd, revised edition, 2006.

\bibitem{GaCo95}
G.~Gallavotti and E.~G.~D. Cohen.
\newblock {Dynamical Ensembles in Nonequilibrium Statistical Mechanics}.
\newblock {\em Physical Review Letters}, 74(14):2694--2697, April 1995,
  arXiv:chao-dyn/9410007.

\bibitem{Gar05}
P.~Garbaczewski.
\newblock {Shannon versus Kullback-Leibler Entropies in Nonequilibrium Random
  Motion}.
\newblock {\em Physics Letters A}, 341(1-4):33--38, June 2005,
  arXiv:cond-mat/0504115.

\bibitem{Gr85}
R.~Graham.
\newblock {Covariant stochastic calculus in the sense of Ito}.
\newblock {\em Physics Letters A}, 109:209--212, 1985.

\bibitem{GuMo83}
F.~Guerra and L.~M. Morato.
\newblock {Quantization of dynamical systems and stochastic control theory}.
\newblock {\em Physical Review D}, 27(8):1774--1786, Apr 1983.

\bibitem{Hsu02}
E.~P. Hsu.
\newblock {\em Stochastic Analysis on Manifolds}, volume~38 of {\em Graduate
  Studies in Mathematics}.
\newblock American Mathematical Society, 2002.

\bibitem{Hsu08}
E.~P. Hsu.
\newblock {A Brief Introduction to Brownian Motion on a Riemannian Manifold}.
\newblock Summer School in Kyushu, 2008.

\bibitem{IkMa79}
N.~Ikeda and S.~Manabe.
\newblock {Integral of Differential Forms along the Path of Diffusion
  Processes}.
\newblock {\em Publications of the Research Institute for Mathematical
  Sciences, Kyoto University}, 15:827--852, 1979.

\bibitem{IkedaWatanabe}
N.~Ikeda and S.~Watanabe.
\newblock {\em Stochastic differential equations and diffusion processes},
  volume~24 of {\em Mathematical Studies}.
\newblock North-Holland, 2 edition, 1989.

\bibitem{It78}
H.~Ito.
\newblock {Probabilistic Construction of Lagrangean of Diffusion Process and
  Its Application}.
\newblock {\em Progress of Theoretical Physics}, 59(3):725--741, 1978.

\bibitem{Ito62}
K.~It\^o.
\newblock {The Brownian Motion and Tensor Fields on Riemannian Manifold}.
\newblock In {\em Proceedings of the International Congress of Mathematicians,
  Stockholm}, pages 536--539. Mittag-Leffler Institute, 1962.

\bibitem{Ko37}
A.~N. Kolmogorov.
\newblock {Zur Umkehrbarkeit der statistischen Naturgesetze}.
\newblock {\em Mathematische Annalen}, 113:766--772, 1937.

\bibitem{KuLe51}
S.~Kullback and R.~Leibler.
\newblock {On Information and Sufficiency}.
\newblock {\em Annals of Mathematical Statistics}, 22(1):79--86, 1951.

\bibitem{Kur98}
J.~Kurchan.
\newblock {Fluctuation theorem for stochastic dynamics}.
\newblock {\em Journal of Physics A: Mathematical and General}, 31(16):3719,
  April 1998,  arXiv:cond-mat/9709304.

\bibitem{Lan61}
R.~Landauer.
\newblock {Irreversibility and heat generation in the computing process}.
\newblock {\em IBM Journal of Research and Development}, 5:183--191, 1961.

\bibitem{LeSp99}
J.~L. Lebowitz and H.~Spohn.
\newblock {A Gallavotti-Cohen Type Symmetry in the Large Deviation Functional
  for Stochastic Dynamics}.
\newblock {\em Journal of Statistical Physics}, 95(1):333--365, March 1999,
  arXiv:cond-mat/9811220.

\bibitem{MaNeWy08}
C.~Maes, K.~Neto\v{c}n\'y, and B.~Wynants.
\newblock {On and beyond Entropy Production: the Case of Markov Jump
  Processes}.
\newblock {\em Markov Processes and Related Fields}, 14(3):445–464, 2008,
  arXiv:arXiv:0709.4327.

\bibitem{MaReMo00}
C.~Maes, F.~Redig, and A.~V. Moffaert.
\newblock {On the definition of entropy production, via examples}.
\newblock {\em Journal of Mathematical Physics}, 41(3):1528--1554, March 2000.

\bibitem{Mey81}
P.-A. Meyer.
\newblock {G\'eom\'etrie stochastique sans larmes, I.}
\newblock {\em S\'eminaire de probabilit\'es de Strasbourg}, 15:44--102, 1981.

\bibitem{Mey82}
P.-A. Meyer.
\newblock {G\'eom\'etrie diff\'erentielle stochastique, II}.
\newblock {\em S\'eminaire de probabilit\'es de Strasbourg}, 16:165--207, 1982.

\bibitem{MiTh06}
T.~Mikami and M.~Thieullen.
\newblock {Duality Theorem for Stochastic Optimal Control Problem}.
\newblock {\em Stochastic Processes and their Applications}, 116:1815--1835,
  2006.

\bibitem{MoSo07}
R.~Mohayaee and A.~Sobolevskii.
\newblock {The Monge-Amp\`ere-Kantorovich approach to reconstruction in
  cosmology}.
\newblock {\em Physica D: Nonlinear Phenomena}, 237:2145--2150, December 2008,
  arXiv:0712.2561.

\bibitem{MG03}
P.~Muratore-Ginanneschi.
\newblock {Path integration over closed loops and Gutzwiller's trace formula}.
\newblock {\em Physics Reports}, 383:299--397, 2003,  arXiv:nlin/0210047.

\bibitem{MGMePe12}
P.~Muratore-Ginanneschi, C.~Mej\'ia-Monasterio, and L.~Peliti.
\newblock {Heat release by controlled continuous-time Markov jump processes}.
\newblock {\em ArXiv Nonlinear Sciences e-prints}, 2012,  arXiv:1203.4062.

\bibitem{Nelson85}
E.~Nelson.
\newblock {\em Quantum fluctuations}.
\newblock Princeton series in physics. Princeton University Press, 1985.

\bibitem{Nelson01}
E.~Nelson.
\newblock {\em Dynamical Theories of Brownian Motion}.
\newblock Princeton University Press, second edition edition, 2001.

\bibitem{Nor92}
J.~R. Norris.
\newblock {A complete differential formalism for stochastic calculus in
  manifolds}.
\newblock {\em S\'eminaire de probabilit\'es de Strasbourg}, 26:189--209, 1992.

\bibitem{Rit08}
F.~Ritort.
\newblock {Nonequilibrium Fluctuations in Small Systems: From Physics to
  Biology}.
\newblock In S.~A. Rice, editor, {\em Advances in Chemical Physics}, volume
  137. John Wiley \& Sons,, April 2008.

\bibitem{ScSe07}
T.~Schmiedl and U.~Seifert.
\newblock {Optimal Finite-Time Processes In Stochastic Thermodynamics}.
\newblock {\em Physical Review Letters}, 98(10):108301, March 2007,
  arXiv:cond-mat/0701554.

\bibitem{Sc31}
E.~Schr\"odinger.
\newblock {\"Uber die Umkehrung der Naturgesetze}.
\newblock {\em Sitzungsberichte der preussischen Akademie der Wissenschaften,
  physikalische mathematische Klasse}, 8(9):144--153, 1931.

\bibitem{Sei05}
U.~Seifert.
\newblock {Entropy Production along a Stochastic Trajectory and an Integral
  Fluctuation Theorem}.
\newblock {\em Physical Review Letters}, 040602(4):95, July 2005,
  arXiv:cond-mat/0503686.

\bibitem{Se98}
K.~Sekimoto.
\newblock {Langevin Equation and Thermodynamics}.
\newblock {\em Progress of Theoretical Physics Supplement}, 130:17--27, 1998.

\bibitem{SiCr12}
D.~A. Sivak and G.~E. Crooks.
\newblock {Thermodynamic Metrics and Optimal Paths}.
\newblock {\em Phys. Rev. Lett.}, 108(19):190602, May 2012,  arXiv:1201.4166.

\bibitem{Topping}
P.~Topping.
\newblock {\em Lectures on the Ricci flow}, volume 325 of {\em London
  Mathematical Society}.
\newblock Cambridge University Press, 2006.

\bibitem{vHa07}
R.~van Handel.
\newblock Stochastic calculus and stochastic control.
\newblock Lecture Notes, Caltech, 2007.

\end{thebibliography}
\bibliographystyle{myhabbrv} %base bibstyle abbrv.bst with eprint field

\end{document}